\documentclass[12pt,twoside]{article}
\usepackage{fleqn,espcrc1,epsfig}

\usepackage{graphicx}
\usepackage[figuresright]{rotating}

\newcommand{\AmS}{{\protect\the\textfont2
  A\kern-.1667em\lower.5ex\hbox{M}\kern-.125emS}}

\title{Probing nucleon strangeness in $\phi$ electroproduction}

\author{Yongseok Oh,%
\address{Center for Theoretical Physics, Seoul National University,
         Seoul 151-742, Korea
         and Institute of Physics, Academia Sinica, Nankang,
         Taipei 11529, Taiwan}%
        \thanks{Address after Aug. 1, 2000: Institute of Physics and
Applied Physics, Yonsei University, Seoul 120-749, Korea}
        Alexander I. Titov,%
\address{Bogoliubov Laboratory of Theoretical Physics, JINR,
               141980 Dubna, Russia}
        Shin Nan Yang\,%
\address{Department of Physics, National Taiwan University,
              Taipei 10617, Taiwan}
and Toshiyuki Morii\,%
\address{Faculty of Human Development, Kobe University,
             Kobe 657-8501, Japan}
}
       
\begin{document}

\maketitle

\begin{abstract}
We investigate $\phi$ meson electroproduction to probe the hidden
strangeness content of the nucleon. 
We found that even a small amount of the $s\bar{s}$ admixture in
the nucleon wavefunction can lead to a significant change in several
double polarization asymmetries in $\phi$ electroproduction, which
can be tested experimentally at current electron facilities.
\end{abstract}

\bigskip

The subject of the strangeness content of the nucleon and nuclei has
been one of the important issues in hadron and nuclear physics
\cite{Ellis00}.
Production of $\phi$ meson from nucleon target has been suggested as
a sensitive probe to study the strangeness content of the nucleon.
Because of the ideal mixing with $\omega$,  $\phi$ meson is nearly a
pure $s\bar{s}$ state and its direct coupling to the nucleon must be
suppressed by the OZI rule.
A non-vanishing $s\bar{s}$ admixture in the nucleon would then give
rise to OZI-evading processes in $\phi$ production and the study of
such processes is expected to provide information on the hidden
strangeness content of the nucleon.
Large deviation from the OZI prediction on $\phi$ production in
$p\bar{p}$ annihilation \cite{AMSL98} has been suggested as a signal of
non-vanishing strangeness content of the nucleon \cite{EKKS95}.
(Different interpretation of such a deviation can be found, e.g., in
Ref. \cite{LZL93}.)

Among the suggested processes, here we consider electromagnetic
production of $\phi$ meson from the nucleon target \cite{HKW92}.
For this process, direct knockout process for $\phi$ production
is allowed if there exits a non-vanishing $s\bar{s}$ admixture in the
nucleon wavefunction.
We have shown that \cite{TOY97-TOYM98}, in the case of $\phi$
photoproduction, knockout process cannot be distinguished from the
background production mechanisms in the differential cross section
measurement.
However, some of the polarization observables were found to be very
sensitive to the strangeness content of the nucleon.
It arises from the interference between the dominant background
production amplitude and the knockout amplitude.
In this work, we extend our previous studies to electroproduction,
i.e., $e+p \to e+p+\phi$, and find similar conclusion to hold.
Namely, some of the double polarization observables of $\phi$
electroproduction can also provide useful tool to probe the hidden
strangeness of the nucleon \cite{TYO97,OTYM99}.

For background processes, we include the diffractive process due to the
Pomeron exchange and one-boson-exchange (OBE) contributions.
The Pomeron exchange process accounts for most of the cross section.
We use the Donnachie-Landshoff model \cite{DL84} to describe the spin
structure of the Pomeron exchange amplitude.
The parameterization of Ref. \cite{FS69} for this amplitude is used
since it gives a good description of the differential cross section in the
considered energy region.
The OBE processes are calculated with the usual effective Lagrangian for
$\pi NN$ ($\eta NN$) and $\pi \gamma \phi$ ($\eta\gamma\phi$) interactions,
together with the Gell-Man--Sharp--Wagner model for the $\phi$ decays,
i.e., $\phi$ decays first into $\rho\pi$ ($\omega\eta$), before converting
into $\gamma\pi$ ($\gamma\eta$) by the vector-meson dominance model.
They give comparable or even larger contributions than that of the Pomeron
exchange in large $|t|$ region.
The parameters for the background processes (Pomeron exchange and OBE)
can be found in Ref. \cite{OTYM99}.

To calculate the direct knockout amplitude, we approximate the proton
wavefunction in Fock space as,
\begin{eqnarray}
|p\rangle = A | [uud]^{1/2} \rangle +
\sum_{j_{s\bar s} = 0,1; \, j_c} b_{j_{s\bar s}}
| [ [ [uud]^{j_{uud}} \otimes
[{\bf L}]]^{j_c}
\otimes [s\bar s]^{j_{s\bar s}} ]^{1/2}
\rangle ,
\label{protonwf}
\end{eqnarray}
where the superscripts $j_{uud}$ ($=1/2$) and $j_{s\bar s}$ ($=0,1$)
denote the spin of the corresponding cluster and $(b_0, b_1)$ correspond
to the amplitudes of the $s\bar s$ cluster with spin $0$ and $1$,
respectively.
The nucleon strangeness content $|B|^2$ is then given by $|B|^2 = |b_0|^2 +
|b_1|^2$ with normalization $|A|^2+|B|^2=1$.
Positive parity of the nucleon ground state requires that the orbital
angular momentum between the clusters be odd and we will restrict
ourselves to $\ell = 1$.
Details of the proton wavefunction is given in  Refs.
\cite{TOY97-TOYM98,TYO97}.

The knockout processes can be classified into $s\bar{s}$-knockout and
$uud$-knockout according to the struck quark cluster.
The contribution of the $s\bar{s}$-knockout is important in the forward
angles, while the $uud$-knockout contributes only at backward angles.
Since the background production mechanisms at large scattering angles
are still not well understood, we focus on the forward scattering angles.
The importance of the intermediate nucleon and nucleon resonances in the
region of large scattering angles has been recently studied
\cite{TLTS99,ZDGS99,CLAS00}.
But the effects of such processes are suppressed at the small scattering
angles and do not affect our conclusion on the forward scattering
angles.

Shown in Fig. 1 are the differential cross sections for each process
as well as the beam-target and beam-recoil double asymmetries with the
longitudinally polarized electron.
We define the beam-target asymmetry $A_{\rm BT}(z)$ with longitudinally
polarized target proton and the beam-recoil asymmetry $A_{\rm BR}(x)$
with transversely polarized recoiled proton by,
\begin{equation}
A_{\rm BT(BR)} =
\frac{d\sigma(\uparrow\downarrow) - d\sigma(\uparrow\uparrow)}
{d\sigma(\uparrow\downarrow) + d\sigma(\uparrow\uparrow)},
\end{equation}
where the arrows represent the helicities of the particles.
We see in Fig. 1 that even if the knockout process is suppressed in
the cross section, its contribution to some of the double polarization
observables is nontrivial because it strongly interferes with the
background production mechanism.
Measuring these asymmetries at the current electron facilities, therefore,
will shed light on our understanding on the hidden strangeness
content of the nucleon.

\begin{figure}[t]
\centerline{\epsfig{file=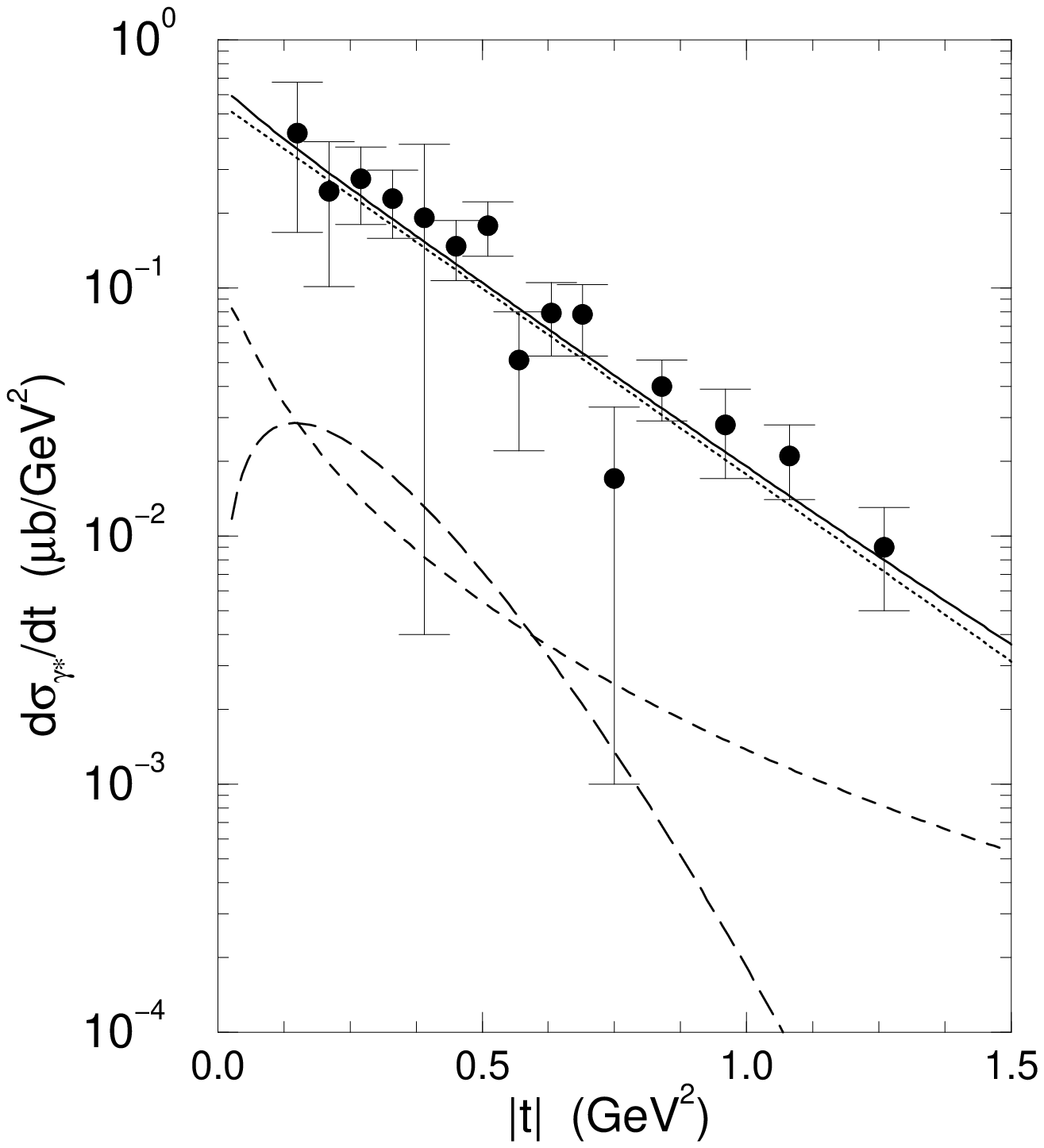, width=5cm} \qquad\qquad
\epsfig{file=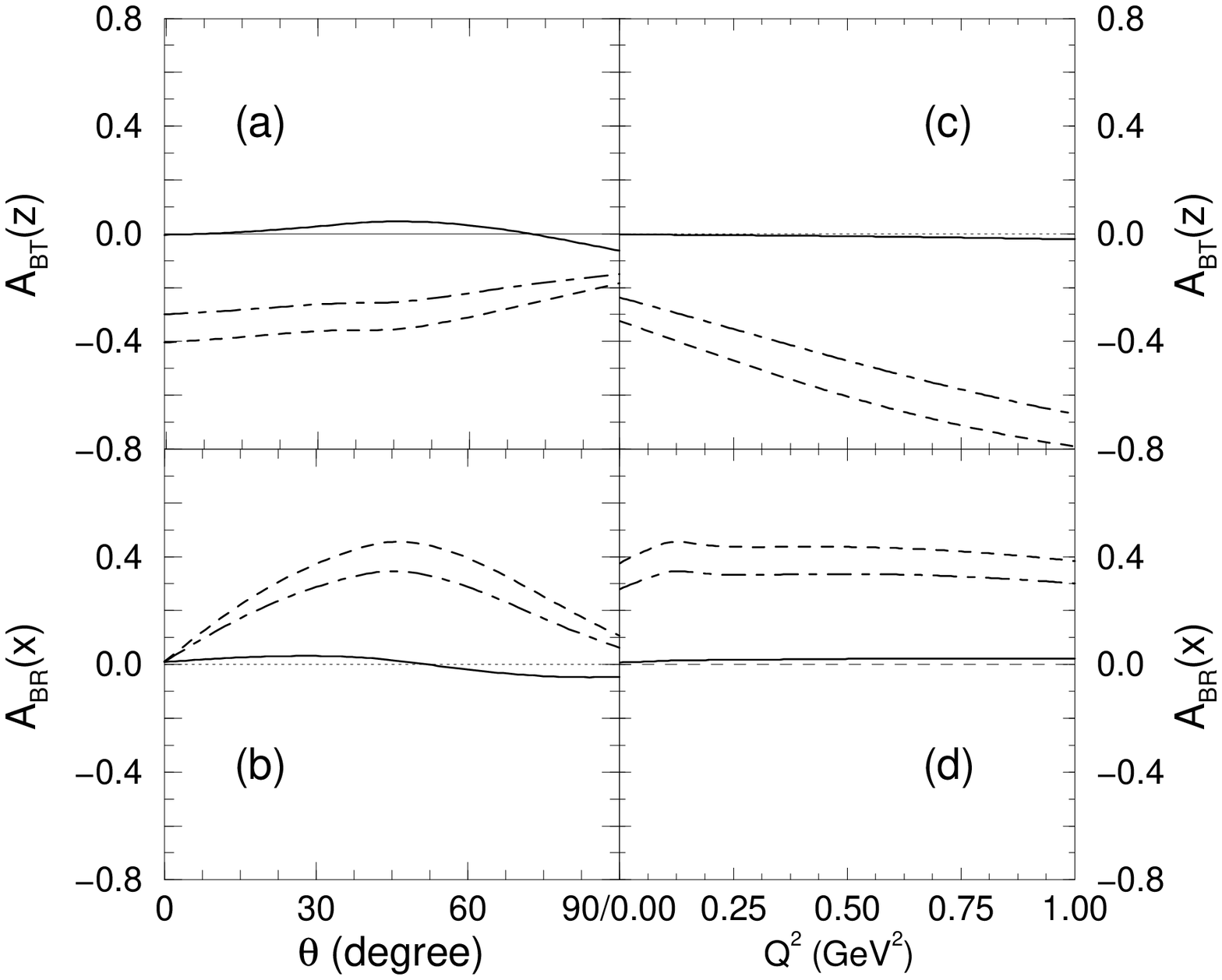, width=7cm}}
\caption{(left panel) The differential cross section for
$\gamma^* p \to \phi p$ at $W = 2.94$ GeV and $Q^2 = 0.23$ GeV$^2$.
The dotted, dashed and long-dashed lines are from Pomeron
exchange, OBE and $s\bar s$-knockout (with $|B|^2 = 1$\%), respectively,
while the solid line shows the Pomeron plus OBE predictions.
The experimental data are from Ref. \cite{DGHH77-DGLS79}.
(right panel) Double polarization asymmetries at $W = 2.15$ GeV.
(a) beam-target double asymmetry at $Q^2 = 0.135$ GeV$^2$ and
(b) beam-recoil double asymmetry at $Q^2 = 0.135$ GeV$^2$ as functions
of the $\phi$ meson production angle $\theta$,
(c) beam-target double asymmetry at $\theta = 0^\circ$ and
(d) beam-recoil double asymmetry at $\theta = 45^\circ$ as functions of
$Q^2$.
The solid, dot-dashed and dashed lines correspond to $|B|^2
= 0$, $0.5$\% and $1.0$\%, respectively.
}
\end{figure}


This work was supported in part by Korea Science and Engineering
Foundation, National Science Council of Republic of China, Russian
Foundation for Basic Research and Monbusho's Special Program for
Promoting Advanced Study (1996, Japan).

\end{document}